\documentclass[journal]{IEEEtran}

\usepackage{multirow}
\usepackage{threeparttable}
\usepackage{booktabs}
\usepackage{amssymb}
\usepackage[noadjust]{cite}
\usepackage{enumerate} 
\usepackage{tabularx,array}
\usepackage{color}
\usepackage[table,xcdraw]{xcolor}
\usepackage{verbatim}
\usepackage[labelsep=period]{caption}
\usepackage{subcaption}

\usepackage{algorithm} 
\usepackage{ulem}
\usepackage{algpseudocode}
\usepackage{hhline}

  \usepackage[pdftex]{graphicx}

  \DeclareGraphicsExtensions{.pdf,.jpeg,.png}

\usepackage[cmex10]{amsmath}

\begin{document}

\title{A Hardware-Efficient ADMM-Based SVM Training  Algorithm for Edge Computing}

\author{
\IEEEauthorblockN{Shuo-An Huang,~\IEEEmembership{Student Member,~IEEE} and Chia-Hsiang Yang,~\IEEEmembership{Senior Member,~IEEE}\\}

\thanks{This work was supported by Ministry of Science and
Technology, Taiwan, under Grant MOST 106-2314-B-002-184. This research was also supported by NOVATEK Fellowship.}

\thanks{S.-A. Huang is with the Graduate Institute of Electronics Engineering, National Taiwan University, Taipei, Taiwan.}

\thanks{C.-H. Yang is with the Department of Electrical Engineering and Graduate Institute of Electronics Engineering, National Taiwan University, Taipei, Taiwan. (e-mail: chyee@ntu.edu.tw).}
}

\maketitle

\begin{abstract}
This work demonstrates a hardware-efficient support vector machine (SVM) training algorithm via the alternative direction method of multipliers (ADMM) optimizer. Low-rank approximation is exploited to reduce the dimension of the kernel matrix by employing the Nystr\"{o}m method. Verified in four datasets, the proposed ADMM-based training algorithm with rank approximation reduces 32$\times$ of matrix dimension with only 2\% drop in inference accuracy. Compared to the conventional sequential minimal optimization (SMO) algorithm, the ADMM-based training algorithm is able to achieve a 9.8$\times$10$^7$ shorter latency for training 2048 samples. Hardware design techniques, including pre-computation and memory sharing, are proposed to reduce the computational complexity by 62\% and the memory usage by 60\%. As a proof of concept, an epileptic seizure detector chip is designed to demonstrate the effectiveness of the proposed hardware-efficient training algorithm. The chip achieves a 153,310$\times$ higher energy efficiency and a 364$\times$ higher throughput-to-area ratio for SVM training than a high-end CPU. This work provides a promising solution for edge devices which require low-power and real-time training. 
\end{abstract}


\IEEEpeerreviewmaketitle

\section{Introduction}
\begin{figure}[t]
\centering
\includegraphics[width=0.41\textwidth]{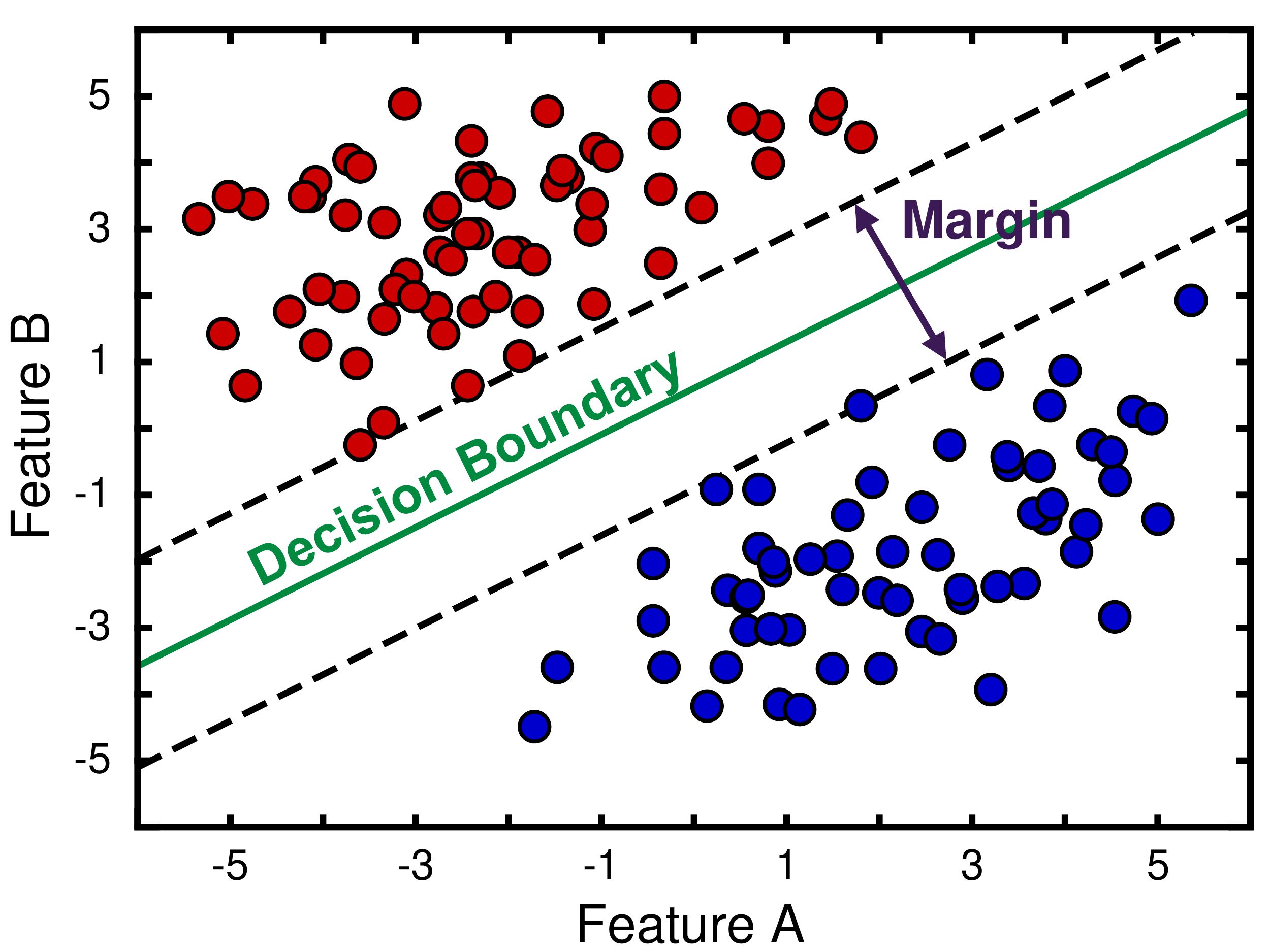}
\caption{A linear SVM classifier finds the decision boundary with the largest margin.} 
\label{linear-svm}
\end{figure}

\IEEEPARstart{I}{n} machine learning, support vector machine (SVM) can solve binary classification problems with a well geometric interpretation \cite{Vapnik:94_svm}. An SVM classifier distinguishes two classes of data by finding a decision boundary from the training samples in the feature space. Among all possible decision boundaries, the SVM seeks the one with the minimum classification error and the maximum decision margin, as shown in Fig. \ref{linear-svm}. Linear SVM was originally devised for binary classification problems \cite{Vapnik:94_svm}. Later on, non-linear SVM \cite{Boser:92_kernelsvm} was proposed to achieve a higher classification accuracy with non-linear decision boundary by applying a kernel transformation. In addition to binary classification, SVM is also utilized to handle multi-class classification problems \cite{Hsu:02_multisvm}. Although neural networks with deep structure have demonstrated excellent classification accuracy in many applications, the SVM is believed to be more effective for classification problems with limited training data \cite{Matykiewicz:12}. Moreover, SVM classifiers feature low computational complexity in inference, which is well suited for edge devices with stringent energy budget. Examples include epileptic seizure detection \cite{Altaf:13_ISSCC}, motion tracking for autopilot \cite{Lee:17_vlsi}, and preliminary object classification and recognition \cite{Jeong:17}. 

For inference, the SVM model is pre-trained and loaded in the devices. However, a well-trained model may not be possible when input features vary over time. For instance, seizure patterns change over time, a pre-trained model \cite{Altaf:13_ISSCC} could become obsolete for accurate long-term detection. Under this situation, model adaptation can be applied for the classification model. In \cite{Lee:13_JSSC}, the recorded data are transmitted to an external device wirelessly and the updated model was sent back for model adaptation. However, the off-line model adaptation poses several issues, including potential data loss and privacy leakage \cite{Xu:14_iot}. Therefore, SVM model adaptation through on-line training is essential \cite{Huang:18_VLSI}.

However, on-line SVM training is challenging for hardware realization. It can be formulated as solving an optimization problem to minimize the cost function associated with the classification error and the magnitude of the margin. Several decomposition techniques, such as sequential minimum optimization (SMO) \cite{Platt:99_SMO}, LIBSVM \cite{Chang:11_libsvm} and cutting-plane methods \cite{Joachims:09_cutplane}, have been developed to solve the SVM training problem. Among these techniques, the SMO algorithm is widely used because it iteratively solves a two-variable quadratic programming (QP) problem, rather than directly solving a large-scale QP problem. However, the nested structure of the SMO limits potential acceleration through hardware parallelism, which leads to a long computation latency.

A hardware-efficient alternative direction method of multipliers (ADMM) \cite{Boyd:11_admm} algorithm is exploited in this work to achieve a short training latency. The ADMM-based SVM training algorithm features a less-loop structure, which can be processed in parallel. ADMM has been utilized to solve linear SVM with feature selection \cite{Ye:11_evsvm}. However, applying the ADMM algorithm to non-linear SVM training for hardware implementation is difficult because large-scale matrices are involved for optimization. Low-rank approximation by employing the Nystr\"{o}m method \cite{Drineas:05_nystrom} is adopted in this work to greatly reduce the matrix dimension, yielding feasible hardware mapping. Several datasets are used to verify the functionality of the proposed training algorithm. The contributions of this work include

\begin{itemize}
\item Applying low-rank approximation to the Nystr\"{o}m method reduces the overall computational complexity and the memory storage reduction by 97\% and 87\%, respectively for the training algorithm.

\item Hardware for eigenvalue decomposition is shared for matrix inversion in Nystr\"{o}m method and solving the linear system in iterative update phase, which achieves 37\% of silicon area reduction.

\item {Pre-computation for matrix matrix achieves computational complexity reduction by 22\% for hardware mapping. }

\item {Combining the common terms by employing intermediate variables reduces the computational complexity by 51\%. }

\item Memory sharing for intermediate variables that reduces the memory usage by 60\% for storing training samples.

\item {A seizure detection chip with on-line model adaptation demonstrates the effectiveness of the proposed SVM training algorithm and the hardware-efficient mapping methodology.}

\end{itemize}

The remainder of this paper is organized as follows. Section II briefly reviews inference and training for SVM. Section III describes the ADMM-based training algorithm and rank approximation. Section IV demonstrates the performance of the ADMM-based training algorithm and shows the superiority of the ADMM-based algorithm. Section V shows the proposed design techniques for efficient hardware mapping. An design example for epileptic seizure detection is shown in Section VI. Finally, Section VII concludes this paper.

\section{Preliminaries of SVM Classifier}
\begin{figure}[t]
\centering
\includegraphics[width=0.5\textwidth]{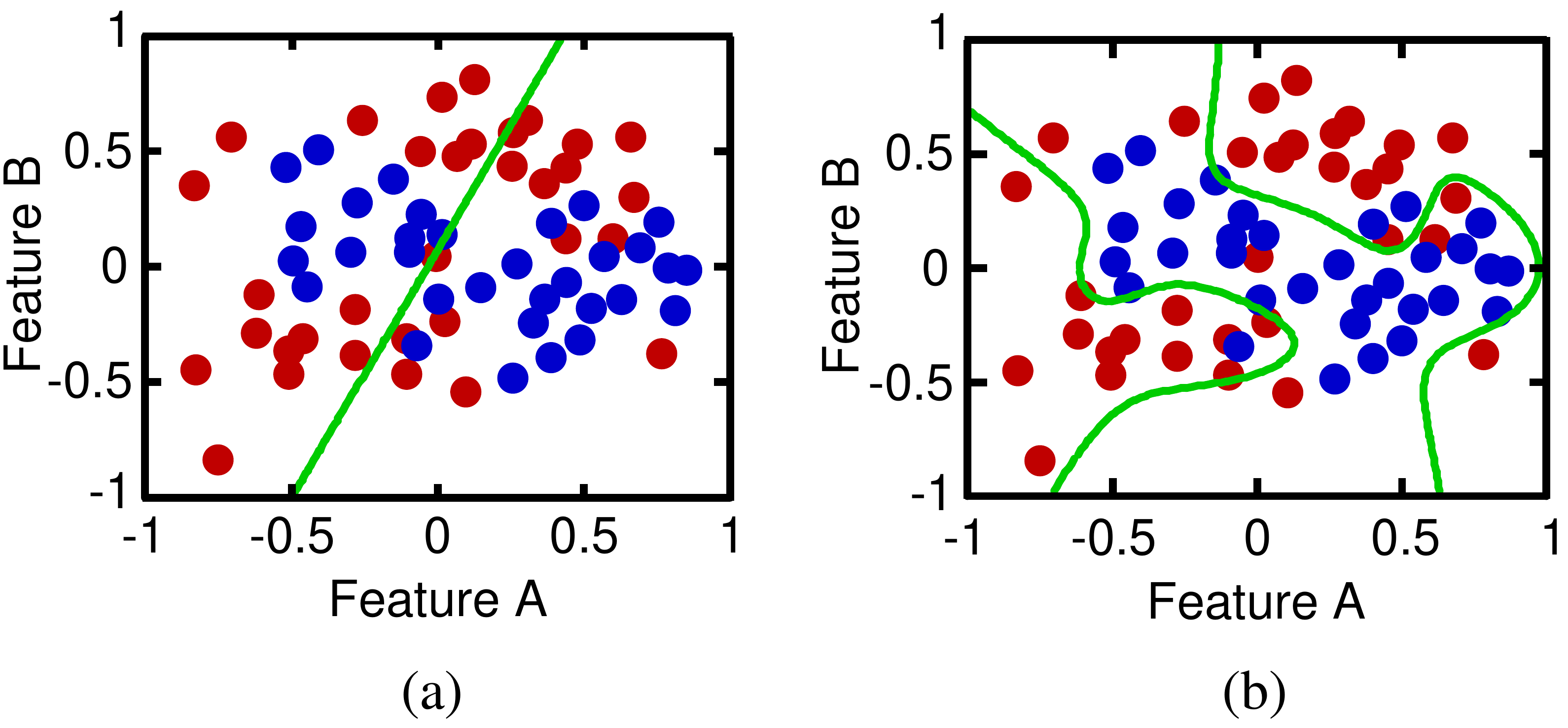}
\caption{Binary classification using (a) linear SVM classifier and (b) non-linear SVM classifier.} 
\label{linear_vs_nonlinear}
\end{figure}

An SVM classifier can be linear or non-linear based on the type of the decision boundary. Fig. \ref{linear_vs_nonlinear} shows both linear and non-linear SVMs. The decision boundary of the linear SVM classifier is an affine hyperplane that separates the data linearly. However, a linear SVM classifier fails to well classify two classes of data when the data are not linearly separable. In contrast, a non-linear SVM classifier improves the classification performance by applying a kernel that transforms the original features into a higher-dimensional feature space \cite{Hofmann:08_kernel}. Generally, the non-linear SVM classifier achieves a better classification performance than the linear one.

\subsection{SVM Inference}
Inference is the prediction process for the data that are not in the training dataset. The SVM inference features low computational complexity. Thus, it is suitable for low-power applications. 

\subsubsection{Linear SVM Inference}
Considering the original $m$-dimensional data $\bf{x}$ and a mapping function $\phi$ that extracts $n$ features from $\bf{x}$ (i.e., $\phi :{\mathbb{R}^m} \to {\mathbb{R}^n}$), the corresponding classification function $\widetilde {{y}}$ can be calculated by
\begin{align} \label{E:1}
\widetilde {{y}} = {\rm{sign}}\left( {  \phi {{\left( {\bf{x}} \right)}^T}{\boldsymbol{\beta }}} + {\beta _0} \right),
\end{align}
where $\boldsymbol{\beta} \in \mathbb{R}^{n}$ is denoted as the model weight and ${\beta _0} \in \mathbb{R}$ is the bias.  

\subsubsection{Non-linear SVM Inference}
A kernel is applied to transform the data from the feature space to a higher-dimensional one where the data can be classified more easily \cite{Hofmann:08_kernel}. Considering $N$ training samples $\left\{ {\left( {{{\bf{x}}_i},{y_i}} \right)} \right\}_{i = 1}^N$, where ${{\bf{x}}_i} \in \mathbb{R}^{m}$ with the associated labels, ${y_i} \in \{  - 1, + 1\} $, and the kernel function ${{k}}\left( {{{\bf{x}}_i},{{\bf{x}}_j}} \right)$, the corresponding classification function $\widetilde {{y}}$ for the unlabelled data $\bf{x}$ can be determined by
\begin{align} \label{E:2}
\widetilde {{y}} = {\rm{sign}}\left( {\mathop \sum \limits_{i = 1}^N {\alpha _i}{y_i}k\left( {{{\bf{x}}_i},{{\bf{x}}}} \right) + b} \right),
\end{align}
where $\boldsymbol{\alpha} \in \mathbb{R}^{N}$ and $b \in \mathbb{R}$ are the model weights and the bias, respectively. A commonly used kernel function is the radial basis function (RBF) (the so-called Gaussian kernel), given by
\begin{align} \label{E:3}
k\left( {{{\bf{x}}_i},{{\bf{x}}_j}} \right) = {\rm{exp}}\left( {\gamma \left\| {{{\bf{x}}_i} - {{\bf{x}}_j}} \right\|_2^2} \right),
\end{align}
where ${\gamma}$ is a negative scaling factor that controls the decay rate of the Gaussian kernel. The RBF demonstrates a non-linear mapping in the Euclidean distance between each training sample ${{\bf{x}}_i}$ and unlabelled data $\bf{x}$. 

Inference for both linear and non-linear SVM classifiers only require multiplications and additions, except that the exponentiation is needed for non-linear SVM inference. The hardware for the exponentiation can be efficiently realized by Coordinate Rotation DIgital Computer (CORDIC) \cite{{Volder:59_CORDIC1}, Meher:09_CORDIC2}. Additionally, only the non-zero model weights and the corresponding features need to be stored, thus SVM classifiers are suitable for edge devices.

\subsection{SVM Training}
The objective of the SVM training is to learn a decision boundary that can separates two classes of data. The SVM model weights associate with a minimized cost that is related to the misclassified error and the model regularization cost. 

\subsubsection{Linear SVM training}
Given the training samples ${{\bf{x}}_i}$ and their corresponding labels $y_i$, SVM training can be formulated as a minimization problem given by
\begin{equation} \label{E:4}
\mathop {\min }\limits_{{\boldsymbol{\beta}} \in {\mathbb{R}^n},{\rm{}}{\beta _0} \in \mathbb{R}} \mathop \sum \limits_{i = 1}^N {\left( {1 - {y_i}\left( {{\beta _0} + \phi {{\left( {{{\bf{x}}_i}} \right)}^T}{\boldsymbol{\beta} }} \right)} \right)_ + } + \frac{\lambda }{2}{||\boldsymbol{\beta}||}_2^2, 
\end{equation}
where { ${\left(  f(\bf{x})  \right)_ + } = {\rm{max}}\left( {1 -  f(\bf{x}) ,0} \right)$} is a Hinge loss function which quantifies the error of mis-classified data \cite{Bishop:06_book}. The regularization term is weighted by a parameter $\lambda$ to control the sparsity of the SVM model weights. The solutions $\boldsymbol{\beta}$ and ${\beta _0}$ in (\ref{E:4}) are the model weights and the bias, respectively, for the linear SVM classifier.

\subsubsection{Non-linear SVM training}
Based on the Karush-Kuhn-Tucker (KKT) condition \cite{KKT}, the model weights $\boldsymbol{\beta}$ of linear SVM classifier can be expressed as a linear combination of the features and their corresponding weights $\boldsymbol{\alpha}$, as represented by 
\begin{align} \label{E:5}
{\boldsymbol{\beta }} = \mathop \sum \limits_{i = 1}^N {\alpha _i}{y_i}\phi \left( {{{\bf{x}}_i}} \right),
\end{align}
which builds the connection to transform linear SVM training to non-linear SVM training. Substituting $\boldsymbol{\beta}$ in (\ref{E:5}) for (\ref{E:4}) and defining the kernel function $k\left( {{{\bf{x}}_i},{{\bf{x}}_j}} \right) = \phi {\left( {{{\bf{x}}_i}} \right)^T}\phi \left( {{{\bf{x}}_j}} \right)$, non-linear SVM training can be reformulated by
\begin{align} \label{E:6}
\mathop {\min }\limits_{{\boldsymbol{\alpha }} \in {\mathbb{R}^N},b \in \mathbb{R}} \mathop \sum \limits_{i = 1}^N {\left( {1 - \left( {{{\boldsymbol{\Psi }}_{i \cdot }}{\boldsymbol{\alpha }} + {y_i}b} \right)} \right)_ + } + \frac{\lambda }{2}{{\boldsymbol{\alpha }}^T}{\boldsymbol{\Psi \alpha }}, 
\end{align}
where ${\boldsymbol{\Psi }} \in {\mathbb{R}^{N \times N}}$ is the kernel matrix which measures the similarities among all the training samples, and each element in ${\boldsymbol{\Psi }}$ is defined by
\begin{align} \label{E:7}
{{{\Psi }}_{ij}}: = {y_i}{y_j}{{k}}\left( {{{\bf{x}}_i},{{\bf{x}}_j}} \right), i, j = 1, 2, 3, ..., N.
\end{align}

For both linear and non-linear SVM classifiers, the computational complexity for the training is much higher than that for inference. A training algorithm that can be efficiently mapped on hardware is required for edge devices.

\section{ADMM-based Training Algorithm}

A hardware-friendly ADMM algorithm is considered in this work. It partitions a large optimization problem into several smaller sub-problems that are easier to solve \cite{Boyd:11_admm}. The ADMM-based SVM training algorithm allows for parallel processing in each iteration, reducing the processing latency. 

\subsection{ADMM-Based Training Algorithm for Linear SVM}

\begin{algorithm*}[t]
\caption{ADMM-based Linear SVM Training Algorithm} \label{algorithm:1}
\begin{algorithmic}[1]
\Require
${\bf{X}} \in \mathbb{R}^{N \times p}$: training samples matrix; 
${\bf{Y}} \in \mathbb{R}^{N \times N}$: diagonal matrix with labels;
\Ensure 
  ${\boldsymbol{\beta}} \in \mathbb{R}^{p \times 1}$: model weight; 
  ${\beta _0} \in \mathbb{R}$: bias;
\State {\bf{initialize}}: ${{\boldsymbol{\beta }}^{\left( 0 \right)}},\beta _0^{\left( 0 \right)}$
\Repeat
  \State Solve linear equation system \hfill \break 
$\left( {\begin{array}{*{20}{c}}
{\lambda {\bf{I}} + {\mu _1}{{\bf{X}}^T}{\bf{X}}}&{{\mu _1}{{\bf{X}}^T}{\bf{1}}}\\
{{\mu _1}{{\bf{1}}^T}{\bf{X}}}&{{\mu _1}n}
\end{array}} \right)\left( {\begin{array}{*{20}{c}}
{{{\boldsymbol{\beta }}^{(k + 1)}}}\\
{{\beta _0}^{(k + 1)}}
\end{array}} \right) = $
$\left( {\begin{array}{*{20}{c}}
{{{\bf{X}}^T}{\bf{Y}}{{\bf{u}}^{(k)}} - {\mu _1}{{\bf{X}}^T}{\bf{Y}}\left( {{{\bf{a}}^{(k)}} - {\bf{1}}} \right)}\\
{{{\bf{1}}^T}{\bf{Y}}{{\bf{u}}^{(k)}} - {\mu _1}{{\bf{1}}^T}{\bf{Y}}\left( {{{\bf{a}}^{(k)}} - {\bf{1}}} \right)}
\end{array}} \right)$

  \State 
${{\bf{a}}^{(k + 1)}} = {S_{\frac{1}{{{\mu _1}}}}}\left( {{\bf{1}} + \frac{{{{\bf{u}}^{(k)}}}}{{{\mu _1}}} - {\bf{Y}}\left( {{\bf{X}}{{\boldsymbol{\beta }}^{(k + 1)}} + {\beta _0}^{(k + 1)}{\bf{1}}} \right)} \right)$
  \State
${{\bf{u}}^{\left( {k + 1} \right)}} = {{\bf{u}}^{\left( k \right)}} + {\mu _1}\left( {{\bf{1}} - {\bf{Y}}\left( {{\bf{X}}{{\boldsymbol{\beta }}^{\left( {k + 1} \right)}} + {\beta _0}^{\left( {k + 1} \right)}{\bf{1}}} \right) - {{\bf{a}}^{(k + 1)}}} \right)$  
  
\Until
$\left\| {\left( {\begin{array}{*{20}{c}}
{{{\boldsymbol{\beta }}^{(k + 1)}}}\\
{{\beta _0}^{(k + 1)}}
\end{array}} \right) - \left( {\begin{array}{*{20}{c}}
{{{\boldsymbol{\beta }}^{(k)}}}\\
{{\beta _0}^{(k)}}
\end{array}} \right)} \right\|_2^2 \le \varepsilon $

\end{algorithmic}
\end{algorithm*}

\begin{figure}[t]
\centering
\includegraphics[width=0.3\textwidth]{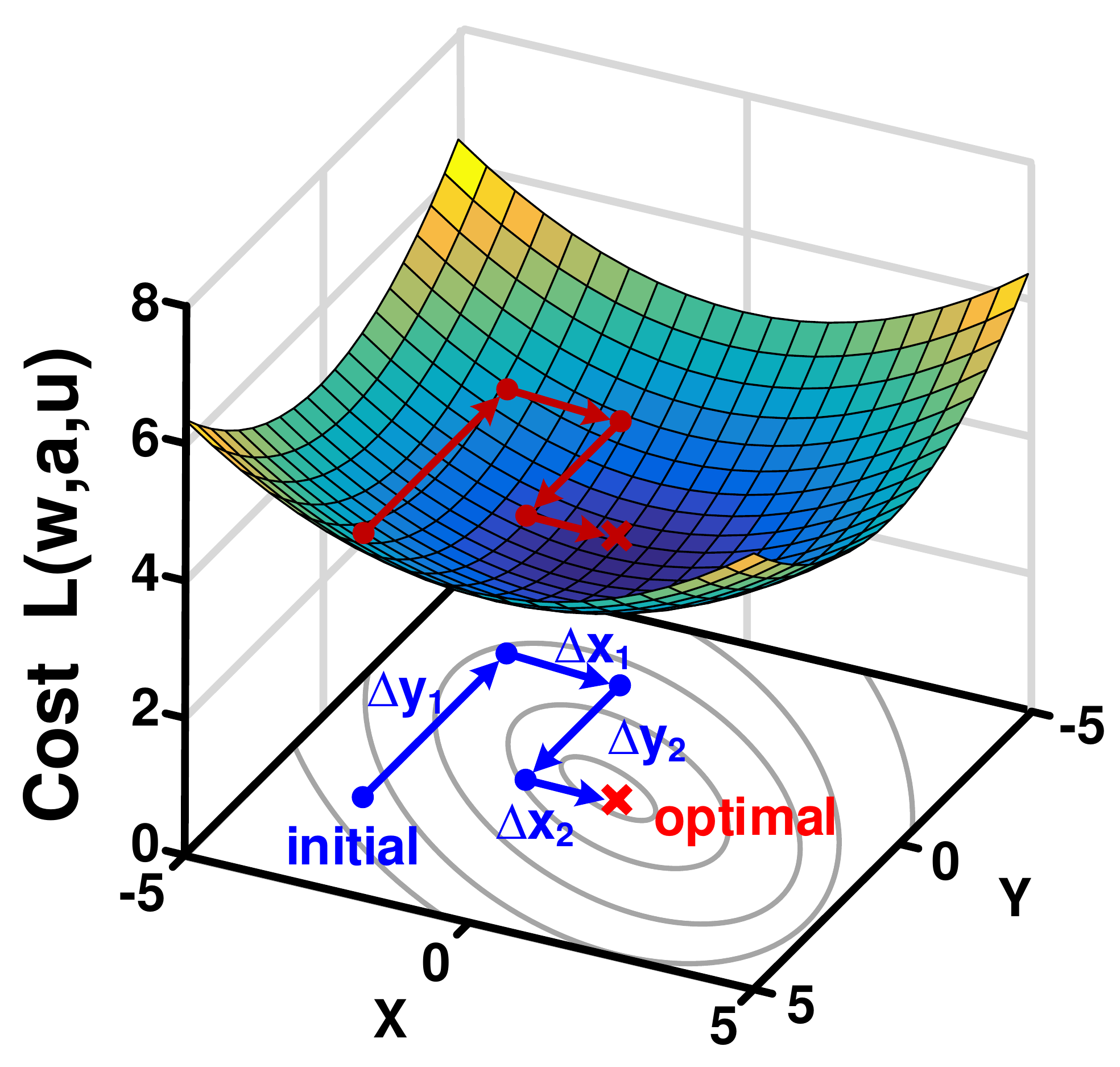}
\caption{The optimization process of ADMM for a two-dimensional convex optimization problem.} 
\label{admm}
\end{figure}

ADMM is exploited to perform variable selection efficiently for SVM training \cite{Ye:11_evsvm}. A similar formulation and optimization procedure are adopted in this work. SVM training in (\ref{E:4}) can be re-written in an equivalent constrained minimization problem in vector form by introducing an auxiliary variable ${\bf{a}} \in {\mathbb{R}^N}$:
\begin{align} \label{E:8}
\mathop {\min }\limits_{{\boldsymbol{\beta }},{\rm{}}{\beta _0}} \mathop \sum \limits_{i = 1}^N {\left( {a_{i}} \right)_ + } + \frac{\lambda }{2}{||\boldsymbol{\beta }||}_2^2, {\quad}s.t.{\quad}{\bf{a}} = {\bf{1}} - {\bf{Y}}\left( {{\bf{X} \boldsymbol{\beta} } + {\beta _0} \bf{1} } \right),
\end{align}
where $\bf{Y}$ is the diagonal matrix with training labels, and $\bf{X}$ is the matrix with training samples, each row of which is a feature vector of a training sample. The augmented Lagrangian function transforms the original problem into an unconstrained convex optimization problem with additional $\ell_{2}$-norm penalty \cite{Ye:11_evsvm}:
\begin{align} \label{E:9}
\begin{split}
L\left( {{\boldsymbol{\beta }},{\beta _0},{\bf{a}},{\bf{u}}} \right) = \mathop {}& \sum \limits_{i = 1}^N {\left( {a_{i}} \right)_ + } + \frac{\lambda }{2}{||\boldsymbol{\beta }||}_2^2 \\
&+ \left\langle {{\bf{u}},{\bf{1}} - {\bf{Y}}\left( {{\bf{X} \boldsymbol{\beta} } + {\beta _0}{\bf{1}}} \right) - {\bf{a}}} \right\rangle \\
{}&+ \frac{{{u_1}}}{2}\left\| {{\bf{1}} - {\bf{Y}}\left( {{\bf{X} \boldsymbol{\beta} } + {\beta _0}{\bf{1}}} \right) - {\bf{a}}} \right\|_2^2,
\end{split}
\end{align}
where $\bf{u}$ is the Lagrange multiplier and ${u_1}$ is the parameter for the penalty term. The ADMM optimizer solves the convex problem in (\ref{E:9}) by splitting it into three sub-problems with respect to variables $\boldsymbol{\beta}$, $\bf{a}$, and $\bf{u}$:
\begin{align} \label{E:10}
\begin{split}
\left( {{{\boldsymbol{\beta }}^{(k + 1)}},{\beta _0}^{(k + 1)}} \right) = {\rm{arg}}\mathop {\min }\limits_{{\boldsymbol{\beta }},{\beta _0}} L\left( {{{\boldsymbol{\beta }}^{(k)}},{\beta _0}^{(k)},{{\bf{a}}^{(k)}},{{\bf{u}}^{(k)}}} \right), \\
{{\bf{a}}^{(k + 1)}} = {\rm{arg}}\mathop {\min }\limits_{\bf{a}} L\left( {{{\boldsymbol{\beta }}^{(k + 1)}},{\beta _0}^{(k + 1)},{{\bf{a}}^{(k)}},{{\bf{u}}^{(k)}}} \right), \\
{{\bf{u}}^{(k + 1)}} = {\rm{arg}}\mathop {\min }\limits_{\bf{u}} L\left( {{{\boldsymbol{\beta }}^{(k + 1)}},{\beta _0}^{(k + 1)},{{\bf{a}}^{(k + 1)}},{{\bf{u}}^{(k)}}} \right),
\end{split}
\end{align}
where $k$ denotes the iteration. In each variable update step, the Lagrangian function $L$ is minimized with respect to a single variable while the other variables remain fixed. Fig. \ref{admm} illustrates the conceptualization of optimization process for a two-dimensional convex problem. The $\boldsymbol{\beta}^{(k)}$ is used to update $\boldsymbol{\beta}^{(k+1)}$, and  $\boldsymbol{\beta}^{(k+1)}$ is then utilized to update $\boldsymbol{\alpha}^{(k)}$, as shown in (\ref{E:10}). Similarly, the new updated variables $\boldsymbol{\beta}^{(k+1)}$ and $\boldsymbol{\alpha}^{(k+1)}$ are used for the update of $\bold{u}^{(k)}$. A variable is updated and the updated variable is then used in the next iteration with respect to another variable. 

{\bf\textsl{Algorithm 1}} shows the ADMM-based linear SVM training algorithm.  The derivative of $L$ with respect to each target variable is set to $0$ to find the closed-form solution for each sub-problem in (\ref{E:10}). A gradient descent method is exploited to minimize the Lagrange function along the negative-gradient direction. The variables are iteratively updated in Steps 3 to 5. The algorithm terminates when the model weight $\boldsymbol{\beta}$ and the bias $\beta _0$ both converge.

Updating ${{\bf{a}}^{(k)}}$, which involves the Hinge loss function, can be solved efficiently. By manipulating the terms in (\ref{E:9}), updating ${{\bf{a}}^{(k)}}$ is reformulated as
\begin{align} \label{E:10-a}
\begin{split}
{{\bf{a}}^{(k + 1)}} &= {\rm{arg}}\mathop {\min }\limits_{\bf{a}} \frac{1}{u_1}\sum \limits_{i = 1}^N {\left( {a^{(k)}_{i}} \right)_ + } \\
&+ \frac{1}{2} \left\|  {{\bf{1}} + \frac{{{{\bf{u}}^{(k)}}}}{{{\mu _1}}} - {\bf{Y}}\left( {{\bf{X}}{{\boldsymbol{\beta }}^{(k + 1)}} + {\beta _0}^{(k + 1)}{\bf{1}}} \right) - {{\bf{a}}^{(k)}} } \right\|_2^2.
\end{split}
\end{align}
The scalar form, which can be applied to vector form element-wisely, is given by

\begin{align} \label{E:10-b}
\begin{split} 
{}&{\rm{Let}} \quad {S_\delta }\left( \theta  \right) = {\rm{arg}}\mathop {\min }\limits_{x } \delta  \cdot {x_ + } + \frac{1}{2}\left\| {x - \theta } \right\|_2^2, \\
{}&{\rm{Then}} \quad {S_\delta }\left( \theta  \right) = \left\{ {\begin{array}{*{20}{c}}
{\theta  - \delta ,}\\
{0,}\\
{\theta ,}
\end{array}} \right.\begin{array}{*{20}{c}}
{\theta  > \delta }\\
{0 \le \theta  \le \delta }\\
{\theta  < 0}
\end{array}.
\end{split}
\end{align}
where $\delta$ is a scalar and $\theta$ is a term that excludes the target variable $x$ (see \cite{Ye:10_proposition} for details). The variable ${{\bf{a}}^{(k + 1)}}$ can be computed efficiently, as described in Step 4 in {\bf\textsl{Algorithm 1}}.

In Step 3, solving the linear equation dominates the overall computational complexity. To solve a linear system  $A\bf{x} = \bf{b}$, a conjugate gradient method \cite{Hestenes:52_conjugate} or a matrix inversion method (i.e., $ {\bf{x}} = A^{-1}\bf{b}$) can be applied. Although the conjugate gradient method avoids matrix inversion and guarantees to solve the linear system within finite iterations, a long latency is usually needed for convergence. By contrary, the matrix inversion method achieves a shorter latency, which is more suitable for real-time applications. Additionally, the system matrix in Step 3 is fixed since the training sample matrix $\bold{X}$ is fixed during the entire training process. Given a fixed system matrix, the inverse of the system matrix can be computed only once and reused for each iteration. This significantly reduces the computational complexity for SVM training.

\subsection{ADMM-Based Training Algorithm for Non-Linear SVM}
The original formulation for non-linear SVM training in (\ref{E:6}) can be solved by the ADMM algorithm and the kernel matrix ${\boldsymbol{\Psi}}$ can be decomposed into several matrices. This transforms a non-linear SVM training problem into a linear SVM training problem with a similar structure \cite{Lee:10_stochastic}. The approximate kernel matrix ${\boldsymbol{\tilde \Psi }}$ can be decomposed by
\begin{align} \label{E:11}
{\boldsymbol{\Psi }} \approx {\boldsymbol{\tilde \Psi }} = {\bf{V}} {{\bf{V}}^T}, 
\end{align}
where $\bf{V} \in {\mathbb{R}^{N \times r}}$ is a rank-$r$ matrix. By substituting ${\bf{V}} {{\bf{V}}^T}$ for ${\boldsymbol{\Psi}}$, (\ref{E:6}) can be reformulated as
\begin{align} \label{E:12}
\mathop {\min }\limits_{{\boldsymbol{\alpha }} \in {\mathbb{R}^N},b \in \mathbb{R}} \mathop \sum \limits_{i = 1}^N {\left( {1 - \left( {{\bf{v}}_i^T{{\bf{V}}^T}{\boldsymbol{\alpha }} + {y_i}b} \right)} \right)_ + } + \frac{\lambda }{2}{{\boldsymbol{\alpha }}^T}{\bf{V}}{{\bf{V}}^T}{\boldsymbol{\alpha }},
\end{align}
where ${\bf{v}_i}$ denotes ${\bf{V}}_{\rm{i}}^{\rm{T}}$, which is defined as the $i$-th row of $\bf{V}$. By replacing ${{\bf{V}}^T}{\boldsymbol{\alpha }}$ with ${\boldsymbol{\eta }}$ and ${\bf{YV}}$ with $\widetilde {\bf{X}}$, (\ref{E:12}) can be rewritten as
\begin{align} \label{E:13}
\mathop {\min }\limits_{{\boldsymbol{\eta }} \in {\mathbb{R}^r},b \in \mathbb{R}} \mathop \sum \limits_{i = 1}^N {\left( {1 - {y_i}\left( {{\bf{\widetilde x}}_i^T{\boldsymbol{\eta }} + b} \right)} \right)_ + } + \frac{\lambda }{2}{{\boldsymbol{\eta }}^T}{\boldsymbol{\eta }}.
\end{align}

Compared with (\ref{E:4}), (\ref{E:13}) is exactly the form of the original problem for a linear SVM training problem. Thus, (\ref{E:13}) can be solved by {\bf\textsl{Algorithm 1}} and the temporary weight coefficient $\boldsymbol{\eta}$ and bias $b$ can be obtained. The original weight coefficient $\boldsymbol{\alpha}$ for the non-linear SVM classifier in (\ref{E:6}) can be recovered from an overdetermined linear system with $N$ unknown variables and $r$ linear equations, given by
\begin{align} \label{E:14}
{\boldsymbol{\eta }} = {{\bf{V}}^T}{\boldsymbol{\alpha }}.
\end{align}
The overdetermined linear system requires at most $r$ non-zero elements for valid solutions. That is, $\boldsymbol{\alpha}$ is a sparse solution, which is an $N \times 1$ vector with $r$ non-zero elements. 

\begin{algorithm}[t]
\caption{ADMM-based non-linear SVM Training Algorithm} \label{algorithm:2}
\begin{algorithmic}[1]
\Require 
  ${\bf{X}} \in \mathbb{R}^{N \times p}$: training samples matrix; 
  ${\bf{Y}} \in \mathbb{R}^{N \times N}$: diagonal matrix with labels;
\Ensure 
  ${\boldsymbol{\alpha}} \in \mathbb{R}^{N \times 1}$: weight vector; 
  $b \in \mathbb{R}$: bias;
\State
${{\boldsymbol{\Psi }}_G} = {\left[ {\left. {{y_i}{y_j}k({x_i},{x_j}} \right)} \right]_{i,j = 1, \cdots N}}$
\State Perform EVD 
${\left[ {{{\boldsymbol{\Psi }}_G}} \right]_{MM}} = {\bf{QD}}{{\bf{Q}}^T}$
\State
${\bf{V}} = {\left[ {{{\boldsymbol{\Psi }}_G}} \right]_{ \cdot ,M}}{{\bf{Q}}_{ \cdot ,r}}{\bf{D}}_{r,r}^{ - 1/2}$
\State
${\bf{X}}^{'} = {\bf{YV}}$
\State
$\left[ {{\bf{w}},b} \right] = {\rm{LinearADMMSolver}}\left( {{{\bf{X}}^{'}},{\bf{Y}}} \right)$ 
\State
${\alpha _i} = \left\{ {\begin{array}{*{20}{c}}
{{{\left[ {{{\bf{Q}}_{ \cdot ,r}}{\bf{D}}_{r,r}^{ - 1/2}{\bf{w}}} \right]}_i}}\\
0
\end{array}\begin{array}{*{20}{c}}
{, \quad i \in M}\\
{, \quad i \notin M}
\end{array}} \right.$

\end{algorithmic}
\end{algorithm}

\subsection{Low-rank Approximation}
Apparently, the sparsity of the ADMM-based SVM model is related to the rank $r$ in (\ref{E:11}), and a dense solution occurs when $\boldsymbol{\Psi}$ is a rank-$N$ matrix. However, a rank-$N$ matrix $\bf{V}$ associated with inversion of an $N \times N$ matrix is infeasible for dedicated hardware mapping when hundreds to thousands of training samples are considered. To reduce the matrix dimension, Nystr\"{o}m method \cite{Drineas:05_nystrom} is exploited in this work to reduce the rank of $\bf{V}$ for approximating the matrix $\boldsymbol{\Psi}$ \cite{Lee:10_stochastic}. Such approximation reduces the dimension for matrix inversion from $ N \times N$ to $r \times r$. 

The Nystr\"{o}m method finds the rank-$r$ matrix $\bold{V}$ for approximating an $N \times N$ symmetric matrix, as given in (\ref{E:11}). The procedure is illustrated as follows. \\

\begin{enumerate}[Step 1:]
\item Specify an integer $c$ with $r \le c \le N$.
\item Randomly choose $c$ elements in set $\{1,2,3,...,N\}$ to generate a subset $M$.
\item Perform eigenvalue decomposition (EVD) on matrix ${{\boldsymbol{\Psi }}_{MM}} = {\bf{QD}}{{\bf{Q}}^T}$, where ${{\boldsymbol{\Psi }}_{MM}}$ is a sub-matrix of $\boldsymbol{\Psi }$ with corresponding row and column indices in subset $M$, ${\bf{Q}} \in \mathbb{R}^{c \times c}$ is an orthogonal matrix, and ${\bf{D}} \in {\mathbb{R}^{c \times c}}$ is a diagonal matrix with decreasing non-negative eigenvalues.
\item Retain the first $r$ entries of $\bf{D}$ as an approximation to produce  ${{\bf{W}}_{c,r}} = {{\bf{Q}}_{ \cdot ,r}}{{\bf{D}}_{r,r}}{{\bf{Q}}_{ \cdot ,r}}^T \in {\mathbb{R}^{c \times c}}$.
\item Perform matrix inversion on ${{\bf{W}}_{c,r}}$ to produce ${\bf{W}}_{c,r}^ +  = {{\bf{Q}}_{ \cdot ,r}}{\bf{D}}_{r,r}^{ - 1}{{\bf{Q}}_{ \cdot ,r}}^T$.
\item Approximated matrix $\widetilde {\boldsymbol{\Psi }} = {{\boldsymbol{\Psi }}_{ \cdot ,M}}{\bf{W}}_{c,r}^ + {{\boldsymbol{\Psi }}_{ \cdot ,M}}^T$. \\
\end{enumerate}

From (\ref{E:11}), Step 4, and Step 6, the matrix $\bf{V}$ can be generated as
\begin{align} \label{E:15}
{\bf{V}} = {{\boldsymbol{\Psi }}_{ \cdot ,M}}{{\bf{Q}}_{ \cdot ,r}}{\bf{D}}_{r,r}^{ - \frac{1}{2}},\quad {\bf{V}} \in {\mathbb{R}^{N \times r}}.
\end{align}

In the Nystr\"{o}m method, the parameters $c$ and $r$ determine the approximation performance and the succeeding classification accuracy. The overall ADMM-based algorithm with rank approximation for non-linear SVM training is summarized in {\bf\textsl{Algorithm 2}}. 

Low-rank approximation reduces the computational complexity for both matrix inversion and variable update. Assume that the rank of the kernel matrix is reduced from $N$ to $r$ by applying the low-rank approximation, the dimension of the input matrix ${\bf{X}}$ in {\bf\textsl{Algorithm 1}} reduces from $N \times N$ to $N \times r$. Compared to the design without low-rank approximation (i.e., ${\bf{X}}$ is full-rank matrix), the computational complexity is approximately reduced by a factor of $N/r$ for each iteration.

\section{Performance Evaluation}
Four datasets: CHB-MIT Scalp Epilepsy \cite{Shoeb:09_MIT, Goldberger:00_chbmit}, Wisconsin Breast Cancer \cite{UCI}, Pima Indians Diabetes \cite{UCI}, and hand-written digit database MNIST \cite{MNIST} are used to verify the effectiveness of the ADMM-based algorithm with rank-approximation. Table \ref{dataset-table} shows the dataset information, including the number of features and the number of data samples for training and inference. Since this work focuses on the binary classification, two classes (digits 4 and 5) in the MNIST database are selected for performance evaluation. In the experiments, the standard deviation for the RBF kernel function $\gamma$ is set to be in the range of [-10, -0.1]. The regularization parameter $\lambda$ and penalty coefficient $u_{1}$ are set to 10 and 1, respectively.

\begin{table}[t]
\centering
\caption{Verification Dataset} 
\includegraphics[width=0.5\textwidth]{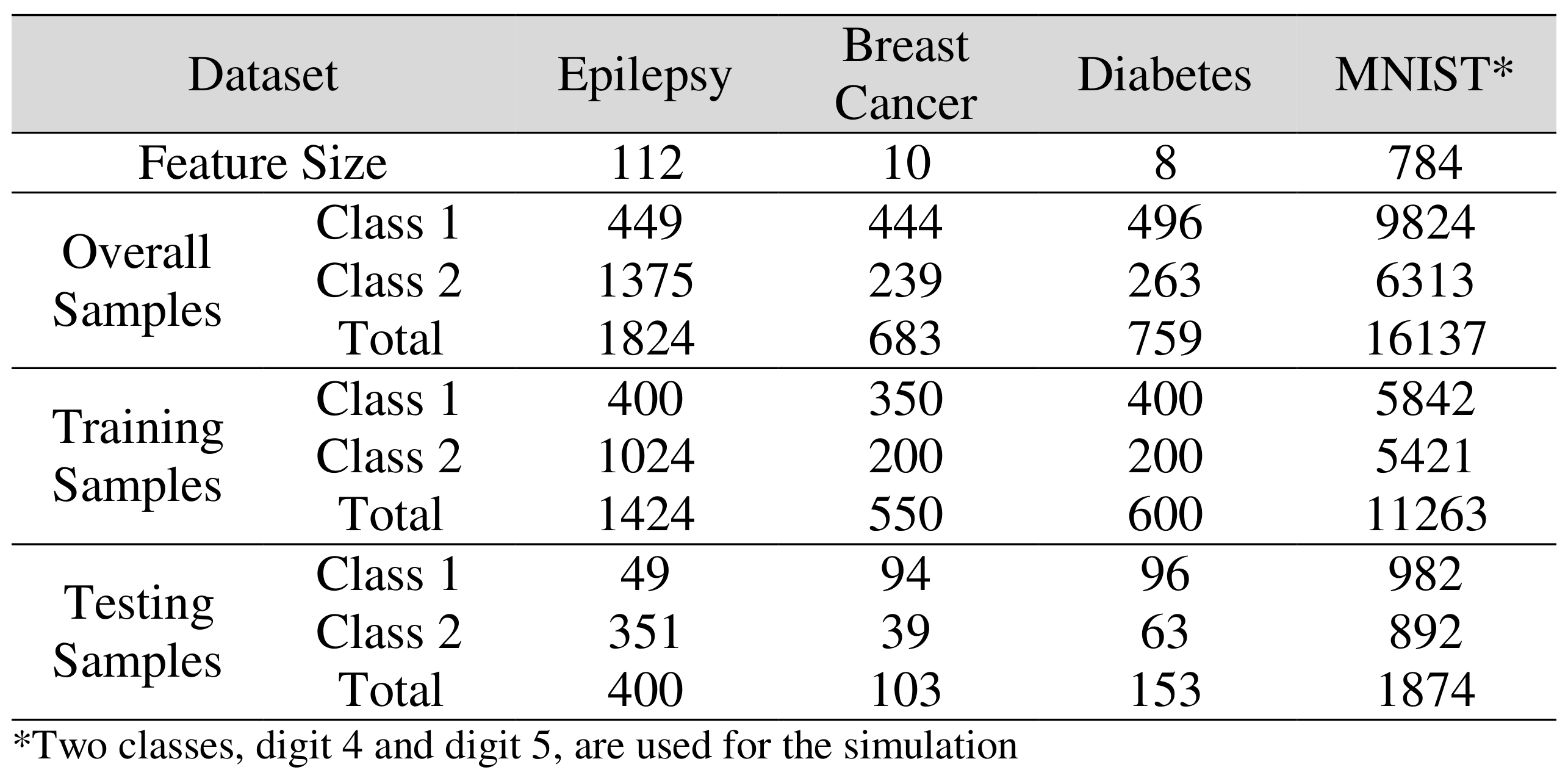}
\label{dataset-table}
\end{table}

\subsection{Settings for Approximated Kernel Matrix}

\begin{figure}[t]
\centering
\includegraphics[width=0.43\textwidth]{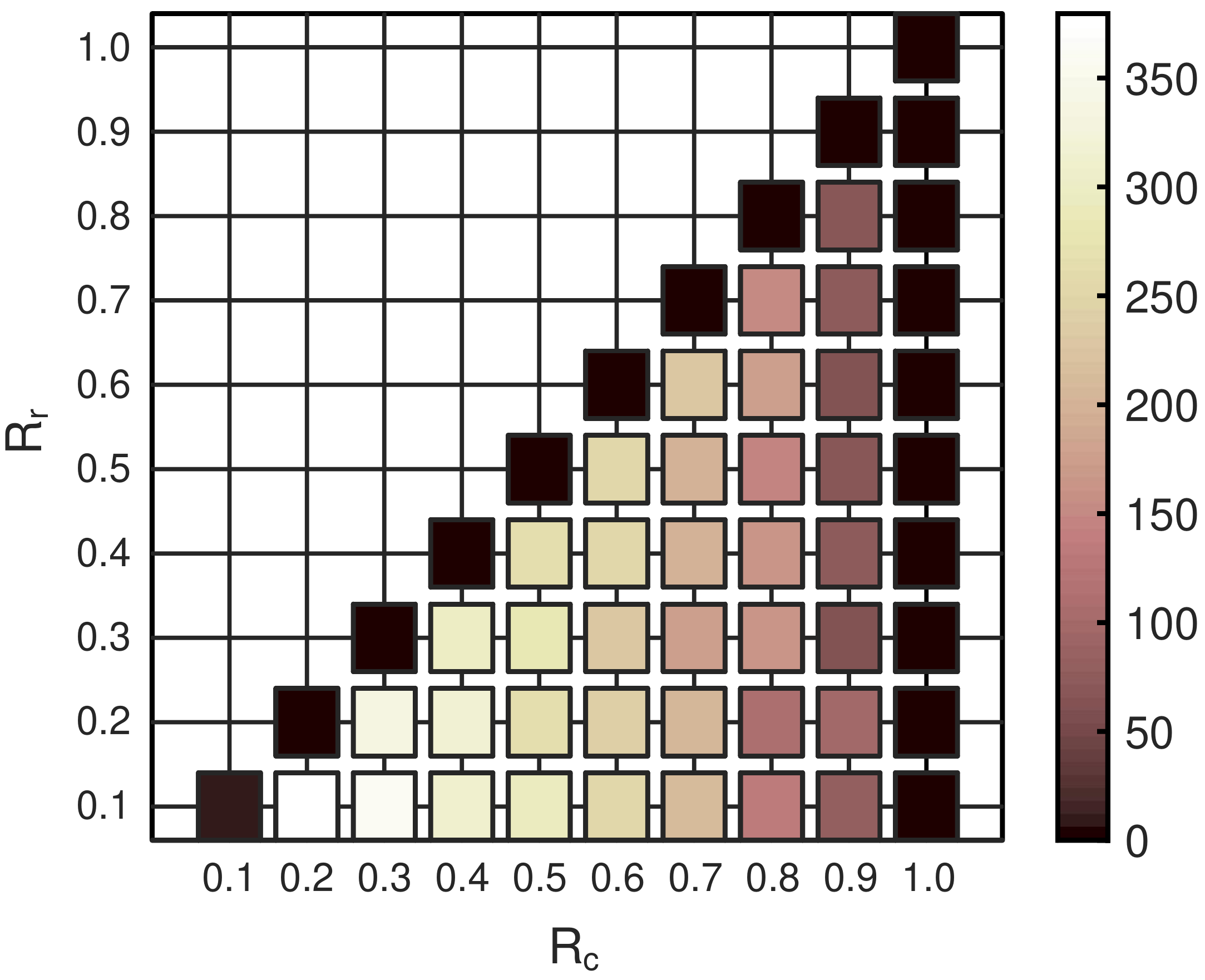}
\caption{MSE analysis for the approximated matrix under various configurations of $(c,r)$. $R_c = c/N$ and $R_r=r/c$, where $N$ is the number of total columns.} 
\label{rank-error}
\end{figure}

In this work, a design methodology is proposed to decide parameters $c$ and $r$ for the Nystr\"{o}m method. The value of $c$ affects both memory storage and computational complexity for implementing the Nystr\"{o}m method and the overall rank reduction is determined by the value of $r$. As an example, the kernel matrix for the MNIST dataset which selects 2048 training samples with 1024 training samples per class is considered. Fig. \ref{rank-error} shows the mean square error (MSE) between the original kernel matrix ${\boldsymbol{\Psi }}$ and the approximated matrix ${\boldsymbol{\tilde \Psi }}$ with respect to parameter pairs $(c, r)$. As can be seen, the approximation error is small for a large value of $c$ because more matrix information is preserved. Additionally, the error can be minimized by setting $c$ to be identical to $r$. Therefore, the design strategy is to choose a small $r$ and to set $c = r$.

\subsection{Classification Accuracy}

\begin{figure}[t]
\centering
\includegraphics[width=0.48\textwidth]{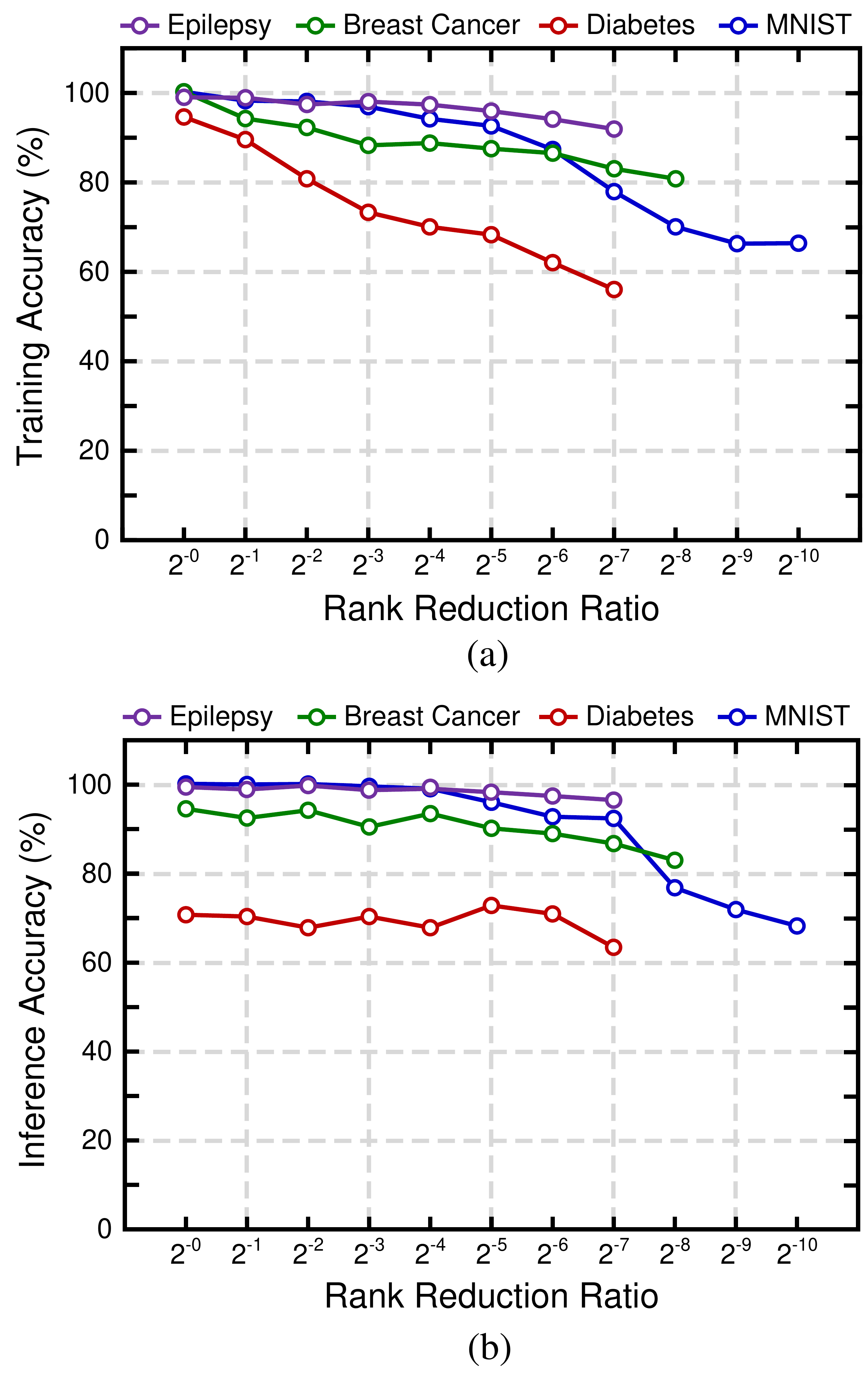}
\caption{The classification accuracy for (a) training and (b) inference, where the rank reduction ratio $R=r/N$.} 
\label{accuracy}
\end{figure}

Fig. \ref{accuracy} shows the classification accuracy for four datasets for training and inference with respect to the rank reduction ratio $R$. For the MNIST dataset, the training accuracy is higher than 95\% even when the rank is reduced to only $2^{-5}$ of the original rank. A 32$\times$ rank reduction is achieved with only a 2\% accuracy drop in inference, which demonstrates the effectiveness of the proposed algorithm. Given the same rank reduction ratio of 2$^{-5}$, only 2.3\% training accuracy and 1.2\% inference accuracy are dropped for the Epilepsy dataset. For the Breast Cancer dataset, a 10.1\% training accuracy drop with 3.2\% inference accuracy drop is achieved. For the Diabetes dataset, the training accuracy decreases significantly, but the inference accuracy remains the same without significant degradation. It is noted that the best classification accuracy for this Diabetes dataset is in the range of 70\% to 76\% \cite{diabetes_result}. In our experiments, the inference accuracy without rank reduction is 70.1\%. By applying the rank approximation, the inference accuracy for Diabetes fluctuates is nearly 70\% and even higher in some cases.

\subsection{Performance Comparison for Rank Approximation}

\begin{figure}[t]
\centering
\includegraphics[width=0.47\textwidth]{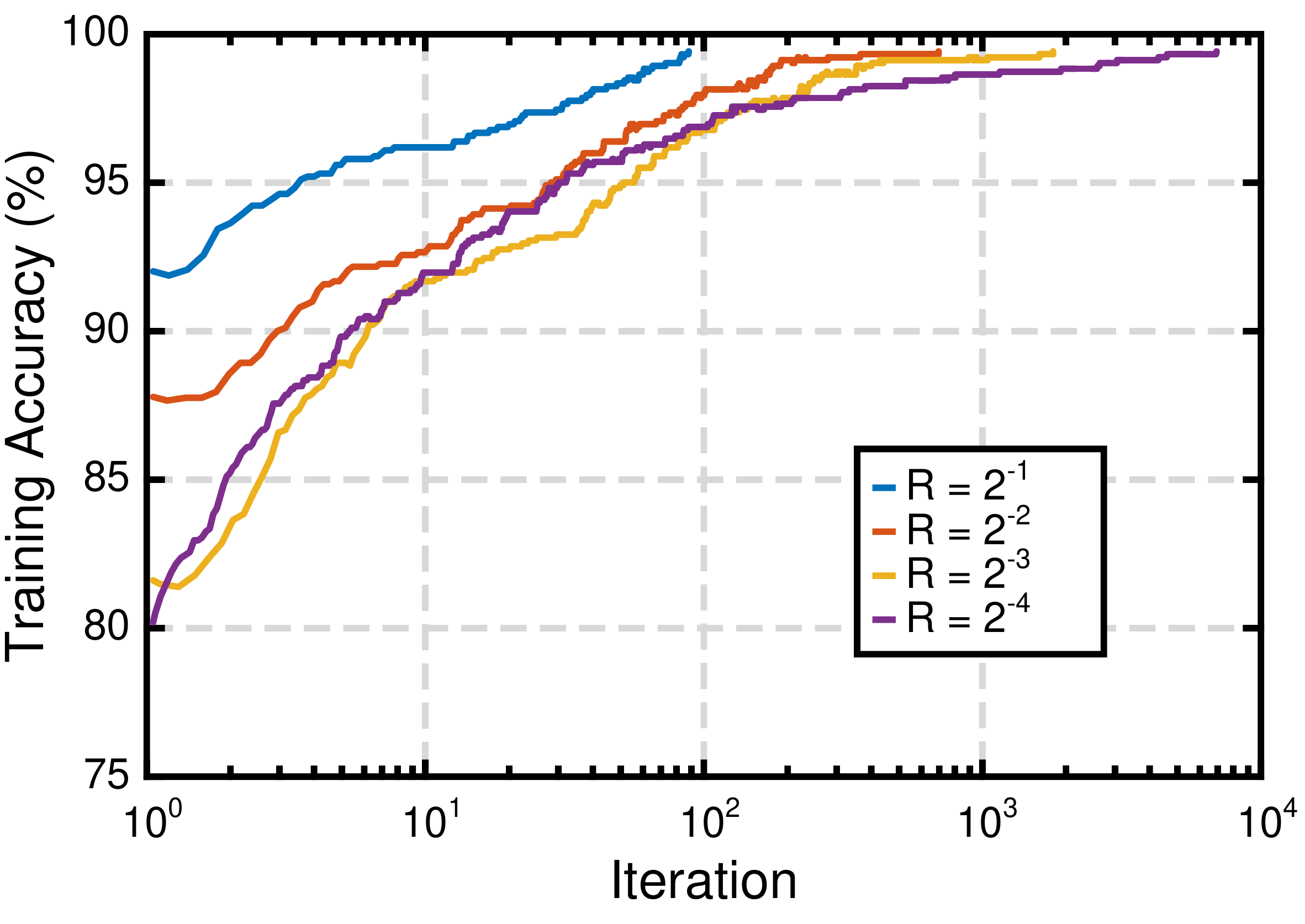}
\caption{The training accuracy of the ADMM-based algorithm versus the iterations with respect to rank reduction ratios $R$.} 
\label{convergent-trend}
\end{figure}

\begin{figure}[t]
\centering
\includegraphics[width=0.45\textwidth]{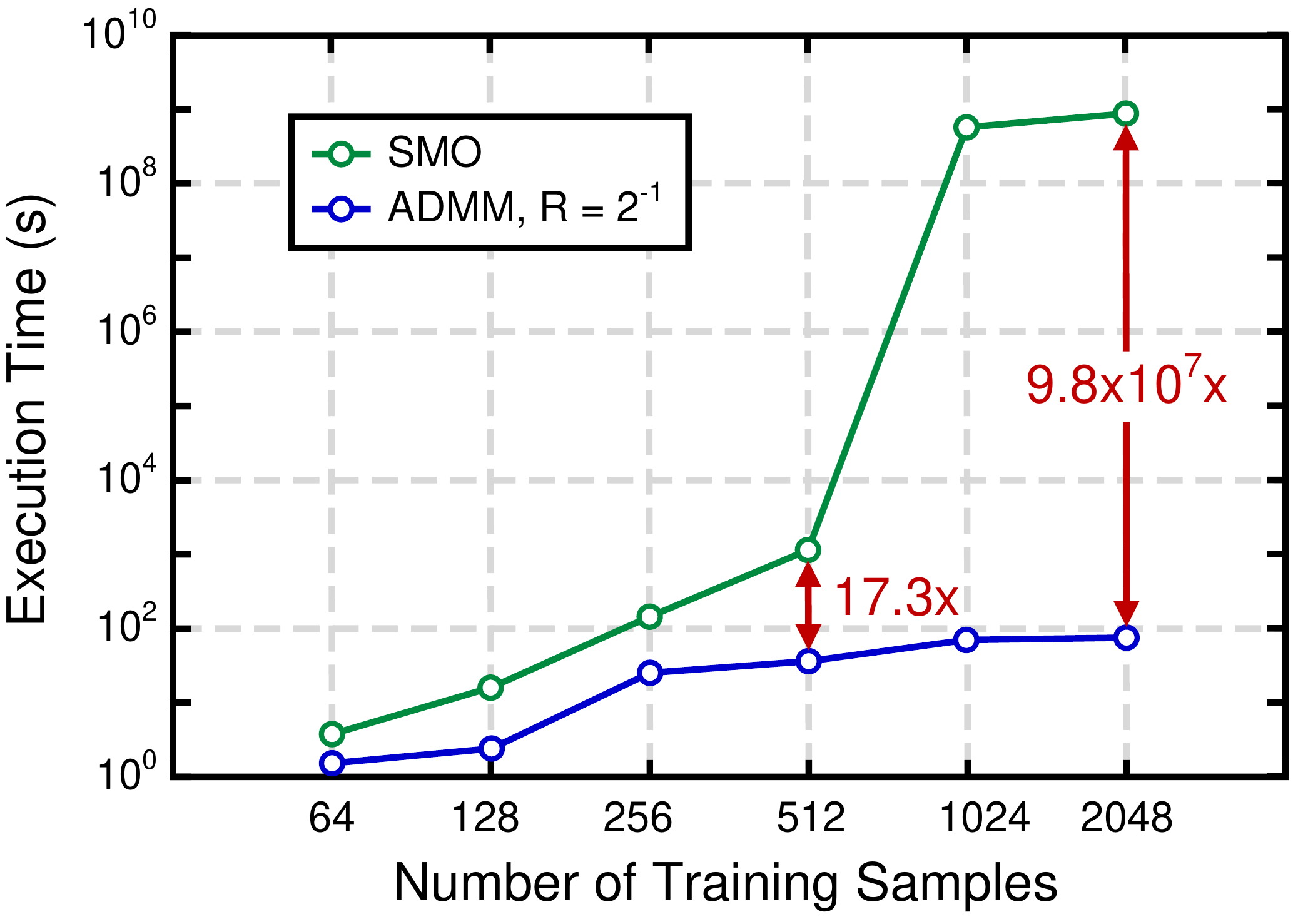}
\caption{The convergence time for the ADMM and SMO algorithms.} 
\label{convergent-time}
\end{figure}

Fig. \ref{convergent-trend} shows the training accuracy versus the iterations with respect to the rank reduction ratio $R$. The training procedure is terminated when a training accuracy of 99\% is achieved. As the number of rank increases, the SVM model initially achieves a better training accuracy and improve its accuracy with less iterations. However, its computational complexity in each iteration also becomes larger than the one of the model with a lower rank. The computational complexity of matrix inversion also increases with number of training samples. Although the model with a smaller rank has a lower initial training accuracy, the training accuracy is improved rapidly to roughly 98\% after similar iterations, and reaches to 99\% gradually. The computational complexity of matrix inversion in each iteration also becomes smaller.

\subsection{Superiority over the SMO Algorithm}

Fig. \ref{convergent-time} shows the time for convergence to reach a training accuracy of 95\% with respect to the number of training samples in the MNIST dataset \cite{MNIST}. The rank reduction ratio $R$ is associated with the model complexity and determines the reduced rank $r$ for rank approximation. The ADMM-based algorithm achieves a 17.3$\times$ faster convergent speed than the SMO algorithm under the case of 512 training samples when simulated on an Intel i7-6700 CPU. The speed improvement becomes larger when the number of training samples increases. The ADMM-based algorithm even overwhelms the SMO algorithm in speed by 9.8$\times$10$^7$ times for the case of 2048 training samples. This is because the SMO algorithm only optimizes two scalar weight coefficients in each iteration. The SMO algorithm fails to update the critical coefficients associated with support vectors since its nested-loop structure. The convergence speed of the ADMM-based algorithm can be further improved through dedicated hardware design.

\section{Efficient Hardware Realization}

Hardware complexity of the ADMM-based algorithm can be further reduced since many operations in {\bf\textsl{Algorithm 1}} can be shared and pre-computed. In this work, several design techniques are proposed for efficient hardware realization.

\subsection{EVD Hardware Sharing}
Since eigenvalue decomposition (EVD) is required in the Nystr\"{o}m method, the matrix inversion in the iterative update step in {\bf\textsl{Algorithm 1}} can be computed by the same EVD hardware. The EVD can be efficiently realized by QR decomposition \cite{Niu:13_QRD} or cyclic Jacobi method \cite{Yang:15_EVD}. The cyclic Jacobi method can be realized by regular rotation operations, which is preferable in hardware implementation. The cyclic Jacobi method performs the EVD by nullifying the off-diagonal elements of the target matrix through a sequence of Givens rotations. It can be greatly accelerated by processing the disjointed rows and columns in parallel.

\begin{algorithm}[t]
\caption{Hardware-efficient ADMM-based Linear SVM Training Algorithm} \label{algorithm:3}
\begin{algorithmic}[1]
\Require 
  ${\bf{X}} \in \mathbb{R}^{N \times p}$: training samples matrix; 
  ${\bf{Y}} \in \mathbb{R}^{N \times N}$: diagonal matrix with training labels;
\Ensure 
  ${\boldsymbol{\beta}} \in \mathbb{R}^{p \times 1}$: weight vector; 
  ${\beta _0} \in \mathbb{R}$: bias;
\State Compute 
${\bf{A}} = \left( {\begin{array}{*{20}{c}}
{\lambda {\bf{I}} + {\mu _1}{{\bf{X}}^T}{\bf{X}}}&{{\mu _1}{{\bf{X}}^T}{\bf{1}}}\\
{{\mu _1}{{\bf{1}}^T}{\bf{X}}}&{{\mu _1}n}
\end{array}} \right)$

\State Perform EVD ${\bf{A}} = {\bf{QD}}{{\bf{Q}}^T}$
\State
${\bf{Z}} = {{\bf{Y}}}\widetilde {\bf{X}}{\bf{Q}}{{\bf{D}}^{ - \frac{1}{2}}}$, where $\widetilde {\bf{X}} = \left[ {\begin{array}{*{20}{c}}
{\bf{X}}&{\bf{1}}
\end{array}} \right]$
\State {\bf{initialize}}: ${{\bf{a}}^{\left( 0 \right)}},{{\bf{u}}^{\left( 0 \right)}}$
\Repeat
  \State 
${\bf{B}} = {{\bf{u}}^{\left( k \right)}} + {u_1}{\bf{1}} - {{\bf{\hat a}}^{\left( k \right)}}$
  \State
${\bf{S}} = {{\bf{Z}}^T}{\bf{B}}$
  \State
${\boldsymbol{\theta }} = {u_1}{\bf{1}} + {{\bf{u}}^{\left( k \right)}} - {u_1}{\bf{ZS}}$
  \State
${{\bf{\hat a}}^{\left( {k + 1} \right)}} = {S_1}\left( {\boldsymbol{\theta }} \right)$
  \State
${{\bf{u}}^{\left( {k + 1} \right)}} = {\boldsymbol{\theta }} - {{\bf{\hat a}}^{\left( {k + 1} \right)}}$
\Until $\left\| {{{\bf{u}}^{\left( {k + 1} \right)}} - {{\bf{u}}^{\left( k \right)}}} \right\| \le \varepsilon $
\State
${\left[ {\begin{array}{*{20}{c}}
{\boldsymbol{\beta }}&{{\beta _0}}
\end{array}} \right]^T} = {\bf{Q}}{{\bf{D}}^{ - \frac{1}{2}}}{\bf{S}}$

\end{algorithmic}
\end{algorithm}

\subsection{Pre-computed Inverse System Matrix}
In Step 3 of {\bf\textsl{Algorithm 1}}, since the training samples are fixed in a single training process, the system matrix ${\bf{A}}$ is also fixed. The inverse of the system matrix can therefore be pre-computed and the precomputed inverse system matrix ${\bf{A}}^{-1}$ can be reused in each iteration. The matrix inversion is computed only once, which greatly reduces the hardware complexity. The matrix inversion, as shown in {\bf\textsl{Algorithm 3}}, is realized by EVD, which can be realized by a shared EVD hardware for the Nystr\"{o}m method, as shown in Step 2 of {\bf\textsl{Algorithm 2}}.

\begin{figure}[t]
\centering
\includegraphics[width=0.5\textwidth]{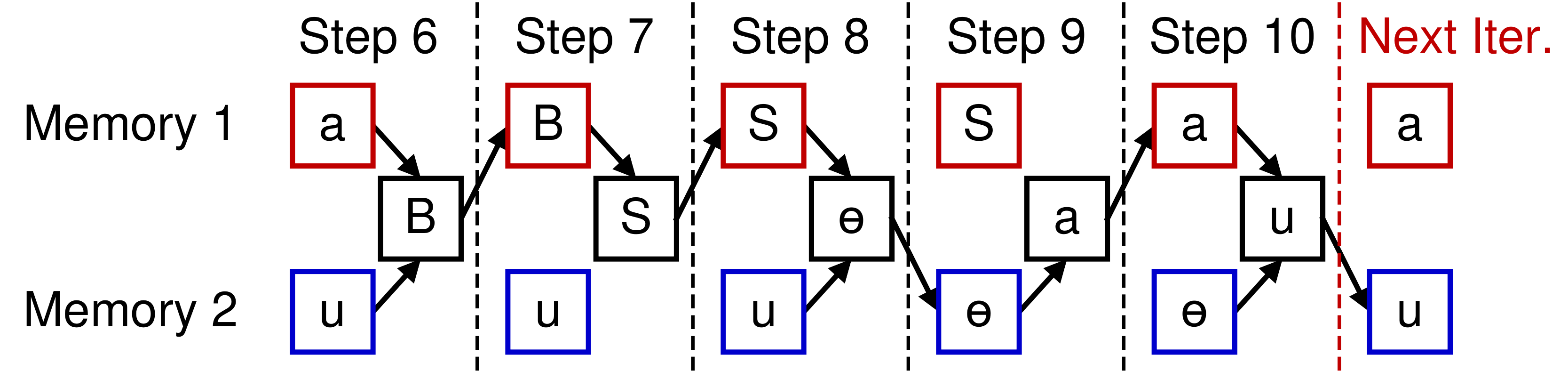}
\caption{Scheduling of memory storage for intermediate variables.} 
\label{reuse-schedule}
\end{figure}

\subsection{Common Term Combination}
Redundant computations in {\bf\textsl{Algorithm 1}} should be avoided to enhance the hardware utilization. The update equations in Step 3 to Step 5 of {\bf\textsl{Algorithm 1}} can be re-written as     

\begin{align} 
\label{E:16}
\widetilde{\boldsymbol{\beta }}^{(k+1)} &= {\bf{A}}^{-1} \widetilde {\bf{X}}^{T}{\bf{Y}} \left( {\bf{u}}^{(k)} + {u_1}{\bf{1}} - {u_1}{\bf{a}}^{(k)} \right), \\ 
\label{E:17}
{\bf{a}}^{(k+1)} &= {S_{\frac{1}{{{\mu _1}}}}} \left( \frac{1}{{u_1}} \left( {u_1}{\bf{1}} + {\bf{u}}^{(k)} - {u_1}{\bf{Y}} \widetilde{\bf{X}} \widetilde{\boldsymbol{\beta }}^{(k+1)}  \right) \right), \\ 
\label{E:18}
{\bf{u}}^{(k+1)} &= \left( {u_1}{\bf{1}} + {\bf{u}}^{(k)} - {u_1}{\bf{Y}} \widetilde{\bf{X}} \widetilde{\boldsymbol{\beta }}^{(k+1)} \right) - {u_1}{\bf{a}}^{(k+1)}, 
\end{align}
where $\widetilde{\boldsymbol{\beta }}$ denotes $\left[ {\begin{array}{*{20}{c}} {\boldsymbol{\beta}}&{\beta_0} \end{array}} \right]^{T}$. By defining a new variable ${\boldsymbol{\theta}}$, given by
\begin{align} \label{E:19}
{\boldsymbol{\theta}} = {u_1}{\bf{1}} + {\bf{u}}^{(k)} - {u_1}{\bf{Y}} \widetilde{\bf{X}} \widetilde{\boldsymbol{\beta }}^{(k+1)},
\end{align}
 (\ref{E:17}) and (\ref{E:18}) can be further simplified as:
\begin{align} 
\label{E:20}
{\bf{a}}^{(k+1)} &= {S_{\frac{1}{{{\mu _1}}}}} \left( \frac{1}{{u_1}} {\boldsymbol{\theta}} \right) = \frac{1}{{u_1}}{S_1 \left( \boldsymbol{\theta} \right) = \frac{1}{{u_1}}} {{\bf{\hat a}}^{\left( {k + 1} \right)}},  \\ 
\label{E:21}
{\bf{u}}^{(k+1)} &= {\boldsymbol{\theta}} - {u_1}{\bf{a}}^{(k+1)} = {\boldsymbol{\theta}} - {{\bf{\hat a}}^{\left( {k + 1} \right)}}, 
\end{align}
as shown in Step 9 and Step 10 of {\bf\textsl{Algorithm 3}}. Note that ${\boldsymbol{\theta}}$ is computed in Step 8 and can be reused in Step 9 and Step 10, which reduces the computational complexity. The minimization solution $S$, which involves Hinge loss function, is reformulated through variable transformation so that the multiplications of $1/u_1$ in (\ref{E:20}) and $u_1$  in (\ref{E:21}) are mutually cancelled out, greatly reducing the hardware complexity.

\begin{figure*}[t]
\centering
\includegraphics[width=0.7\textwidth]{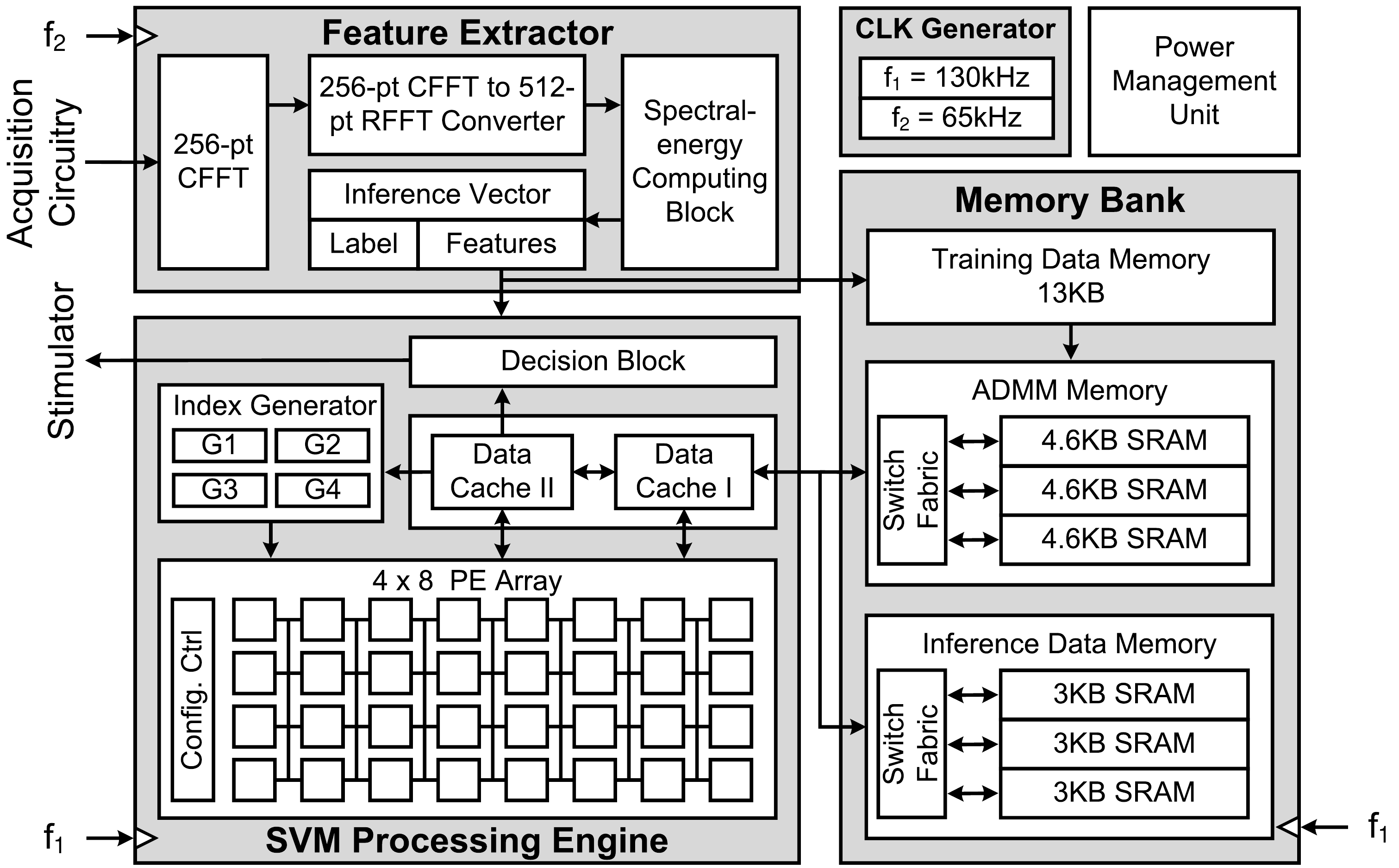}
\caption{Architecture of the ADMM-based SVM processor with on-chip learning for epileptic seizure detection \cite{Huang:18_VLSI}. } 
\label{system-architecture}
\end{figure*}

\subsection{Memory Sharing for Intermediate Variables}
The matrix ${\bf{A}}^{-1}$ remains constant throughout the entire training process. It requires an extra  memory with a size of $(p+1) \times (p+1)$ to store ${\bf{A}}^{-1}$. Instead of introducing extra memory, a memory reuse strategy is proposed to achieve efficient hardware mapping. By substituting $\widetilde{\boldsymbol{\beta }}$ in (\ref{E:16}) for (\ref{E:19}), ${\boldsymbol{\theta}}$ can be computed as:
\begin{align} 
\label{E:22}
{\boldsymbol{\theta}} &= {u_1}{\bf{1}} + {\bf{u}}^{(k)} - {u_1}{\bf{Y}} \widetilde{\bf{X}} {\bf{Q}}{\bf{D}}^{-1}{\bf{Q}}^{T} {\widetilde{\bf{X}}}^{T}{\bf{Y}}^{T}{\bf{B}} \\
\label{E:23}
&= {u_1}{\bf{1}} + {\bf{u}}^{(k)} - {u_1}{\bf{M}}{\bf{B}},
\end{align}
where ${\bf{B}}$ is defined in Step 6 of {\bf\textsl{Algorithm 3}}, and ${\bf{Q}}{\bf{D}}{\bf{Q}}^{T}$ is the result of the eigenvalue decomposition of the matrix ${\bf{A}}$. The matrix ${\bf{M}}$, which combines all the constant terms, is an $N \times N$ matrix. It is impractical to store such a large matrix. By leveraging pre-computation and the memory reuse, a new variable ${\bf{Z}}$, which is an $N \times (p+1)$ matrix, is introduced to store the pre-computed matrix and to reuse the memory space occupied by training sample matrix $\widetilde{\bf{X}}$, which is also an $N \times (p+1)$ matrix, as shown in Step 3 in {\bf\textsl{Algorithm 3}}. Therefore, $\widetilde{\bf{X}}$, which has already merged into $\bf{Z}$, is no longer needed in the subsequent steps. The proposed technique results in a computational complexity reduction without extra memory overhead.

Scheduling for memory storage is further improved for intermediate variables to achieve high memory utilization. The ${\boldsymbol{\theta}}$ can be computed by
\begin{align} 
\label{E:24}
{\boldsymbol{\theta}} &= {u_1}{\bf{1}} + {\bf{u}}^{(k)} - {u_1}{\bf{Z}}{\bf{Z}}^{T}{\bf{B}} \\
&= {u_1}{\bf{1}} + {\bf{u}}^{(k)} - {u_1}{\bf{Z}}{\bf{S}},
\end{align}
where variable ${\bf{S}}$ is introduced to split the computation into two steps, as shown in Step 7 and Step 8 of the {\bf\textsl{Algorithm 3}}. Fig. \ref{reuse-schedule} shows the memory allocation for intermediate variables over time. The intermediate variables $\bf{B}$, $\bf{S}$, ${\boldsymbol{\theta }}$, $\bf{\hat{a}}$, and $\bf{u}$ share two caches and have minimum dependency for few steps. For example, the variable $\bf{B}$ is generated in Step 6, and only be used in Step 7 to compute the variable $\bf{S}$. After Step 7, the variable $\bf{B}$ is no longer required and therefore the occupied memory can be released for other variables, such as variable $\bf{S}$. Only two ($N \times 1$) memory banks are needed for these intermediate variables.

\section{Design Example}
An SVM-based seizure detector with on-chip learning for epilepsy neuro-modulation is demonstrated in \cite{Huang:18_VLSI}. It adopts the proposed ADMM-based training algorithm to achieve high efficiency in both energy and area metrics. Fig. \ref{system-architecture} shows the system architecture of the ADMM-based seizure detector. The architecture includes a FFT-based feature extractor to extract spectral-energy features, SRAM memory banks to store training samples and intermediate data, and an SVM processing engine to handle all the computations that requires in both SVM training and inference phases. 

Low-rank approximation enables a lower matrix dimension for iterative update and the matrix inversion. This greatly reduces the complexity for the hardware mapping. The rank is reduced from 256 to 16, which results in a 97\% computational complexity reduction and an 87\% memory storage reduction compared to the direct-mapped hardware mapping. 

The SVM processing engine includes a CORDIC-based processing element (PE) array, register banks, and an index generator for Jacobi rotation indices.  The CORDIC-based PE is exploited to support all the linear and non-linear operations. The PE array can be configured as a multiply-and-accumulate (MAC) or an adder tree to support different vector manipulations in the iterative update step. Additionally, a systolic PE array is designed to perform high-speed EVD. The hardware for EVD is shared for Nystr\"{o}m method in rank approximation and matrix inversion for solving the linear system in the variable update phase, which reduces the silicon area by 37\%. The approximate Jacobi method \cite{Gotze:93_TC} is adopted to reduce the processing latency. Unlike the cyclic Jacobi method, which requires several cycles for vector rotation and angle generation, the approximate Jacobi method only needs one cycle. The approximate Jacobi method achieves an 8.9$\times$ convergence speed than the conventional cyclic Jacobi method. By rotating the disjointed row and column pairs in parallel, the processing latency is reduced by 8 times. The overall latency reduction for EVD is reduced by 98.6\% compared to the direct-mapped design that realizes the cyclic Jacobi method. 

By decomposing the matrix $\bf{A}$ into $\bf{Q}\bf{D}{\bf{Q}}^{T}$, $\bf{Q}$ and $\bf{D}$ can be combined with $\bf{Y}$ and $\widetilde{\bf{X}}$ to form the matrix $\bf{Z}$. The EVD of $\bf{A}$ dominates the computational complexity of a single update iteration. By this pre-computation scheme, the EVD of $\bf{A}$ is computed only once, and 22\% of the computational complexity is saved for an iteration compared to the design without pre-computation. The computational complexity can be further reduced by 51\% by reusing ${\boldsymbol{\theta }}$ and $\bf{Z}$. The overall computational complexity is reduced by 62\% by exploiting pre-computation and common term combination. Two caches are exploited to store intermediate variables in the iterative update step. Since the variables are alternately used and only active for a short period, the memory banks that store these variables can be shared. The register memory is reduced by 60\% through proper scheduling. Furthermore, since the training sample memory $\bf{X}$ is dispensable in iterative update step, the SRAMs used for storing $\bf{X}$ are fully reused for storing $\bf{Z}$. No extra memory is needed to store the pre-computed $\bf{Z}$.

The seizure detector chip supports both inference and training. It is the first silicon proof of SVM training on chip. The chip can perform SVM training in 0.78 second and only dissipates 2.9 mW. Compared to an Intel i7-2600 CPU, the chip achieves a 153,310$\times$ higher energy efficiency and a 364$\times$ higher throughput-to-area ratio for SVM training.

\section{Conclusions}

SVM is a powerful classifier for applications that lack sufficient training samples and ask for low power dissipation. It features simple computation in inference but the computational complexity for training is still very high. On-line training has drawn attention for model adaptation at the edge devices. Compared to the conventional SMO algorithm, ADMM-based algorithm features a short convergence latency and sparse weights. Low rank approximation via Nystr\"{o}m method is applied for feasible hardware mapping, which greatly reduces the overall computational complexity for SVM training. Four open datasets are used to demonstrate the effectiveness of the ADMM-based algorithm with low rank approximation. To achieve efficient hardware mapping, several hardware techniques are proposed to further reduce the hardware complexity. Pre-computation for inverse matrix and combining common terms jointly reduce the computational complexity by 62\%. Memory sharing for intermediate variables reduces 60\% of memory usage. As a proof of concept, a seizure detector chip showcases the high efficiency of the proposed architecture for SVM training. The chip dissipates less than 3 mW and achieves orders of magnitude improvement in energy efficiency when compared to a high-end CPU. The proposed algorithm allows low-power, real-time SVM training on chip, providing a promising solution for smart edge devices.

\section*{Acknowledgment}
The authors thank the Taiwan Semiconductor Research Institute (TSRI) for technical support on chip design.

\bibliographystyle{IEEEtran}

\vfill

\end{document}